\documentclass[prl,twocolumn,aps]{revtex4}
\usepackage{epsfig}
\usepackage{bm}
\usepackage[dvips]{color}
\usepackage{lscape}
\newcommand{\B}[1]{{\bm{#1}}}

\usepackage[latin1]{inputenc}
\newcommand{\beq}{\begin{equation}}
\newcommand{\eeq}{\end{equation}}
\newcommand{\bea}{\begin{eqnarray}}
\newcommand{\eea}{\end{eqnarray}}
\newcommand{\pa}{\partial}

\begin{document}

\title{Cracklike Dynamics at the Onset of Frictional Sliding}

\author{Eran Bouchbinder$^1$, Efim A. Brener$^{1,2}$, Itay Barel$^3$ and Michael Urbakh$^3$}
\affiliation{$^1$ Chemical Physics Department, Weizmann Institute of Science, Rehovot 76100, Israel\\
$^2$ Peter Gr\"unberg Institut, Forschungszentrum J\"ulich, J\"ulich 52425 Germany\\
$^3$ School of Chemistry, Tel Aviv University, Tel Aviv 69978, Israel}


\begin{abstract}
We propose an elasto-plastic inspired friction model which incorporates interfacial stiffness. Steady state sliding friction is characterized by a generic nonmonotonic behavior, including both velocity weakening and strengthening branches. In 1D and upon the application of sideway loading, we demonstrate the existence of transient cracklike fronts whose velocity is independent of sound speed, which we propose to be analogous to the recently discovered slow interfacial rupture fronts. Most importantly, the properties of these transient inhomogeneously loaded fronts are determined by steady state front solutions at the {\em minimum} of the sliding friction law,
implying the existence of a new velocity scale and a ``forbidden gap'' of rupture velocities. We highlight the role played by interfacial stiffness and supplement our analysis with 2D scaling arguments.
\end{abstract}
\maketitle

The frictional strength and stability of spatially extended interfaces is important for a wide range of natural and man-made systems \cite{00Persson, 01BenZion, 04UGI, 06BC, 08BenZion}. Yet, several fundamental aspects of it are not well understood. One of these aspects is the onset of frictional sliding in which the (initially static) interface separating two bodies in frictional contact fails under shear forces, giving rise to relative shear motion.

This transition was recently observed to be mediated by the propagation of interfacial fronts \cite{04RCF, 07RCF, 10BDRF, 10BDCF}. By directly tracking the real contact area between the two bodies, it was demonstrated that in addition to fast sub-Rayleigh and super-shear cracklike modes \cite{04RCF, 10BDCF, 04XRK}, there exist also slow cracklike modes which travel at well-defined velocities significantly smaller than the Rayleigh wave speed, but nevertheless play a significant role in interface weakening. Furthermore, these experiments highlight the key role of inhomogeneity to frictional stability and mode selection \cite{10BDCF}, and the relevance of plastic deformation of interlocking asperities to frictional strength \cite{10BDRF}. These important and unexplained observations, especially the nature and properties of the slow cracklike fronts, is our main focus here.

In this Letter we develop a dry friction model based on the dynamics of microcontacts at frictional interfaces. We demonstrate the existence of a frictional instability prior to the onset of sliding, which excites cracklike fronts whose velocity is independent of sound speed. Most importantly, we show that the properties of these slow fronts propagating under transient inhomogeneous conditions is determined by steady state front solutions at the minimum of the sliding friction law, where velocity-weakening behavior crosses over to velocity-strengthening one \cite{06BC, 10RL}. Our work is strongly influenced by recent important numerical investigations of a microscopic 1D model \cite{09BBU} and macroscopic 2D analytic results \cite{02BM, 05BMM}.

Our starting point is the following expression for the total frictional stress $\sigma_{xy}(x,y\!=\!0,t)$, where $y\!=\!0$ is the location of the interface and $x$ is the position along it,
\begin{equation}
\label{separation}
\sigma_{xy}(x,y\!=\!0,t) = \bar\sigma(x,t) + \eta \pa_t u_x(x,y\!=\!0,t) \ .
\end{equation}
Here $\bar\sigma$ is associated with interfacial contact dynamics, $\eta$ is a viscous-friction coefficient and $\pa_t u_x$ is the slip rate. We choose the velocity-strengthening part $\eta \pa_t u_x$ to have the simplest possible form, which does not affect the generality of our results, as long as such a branch exists. In addition, we set $u_y(x,y\!=\!0,t)\!=\!0$.

The next step is to write down dynamic equations for $\bar\sigma$ in terms of $A$, the ratio between the real contact area and the nominal one, in the form
\begin{eqnarray}
\label{contact} \pa_t A &=& \frac{A_0 - A}{\tau_0} - \theta\left(\frac{|\sigma_{xy}|}{A}-\sigma_c\right) \frac{\kappa ~ A}{\tau_1} \ ,\\
\label{friction} \pa_t \bar\sigma &=& \frac{\mu_0 \,A \,\pa_t u_x}{h}  - \theta\left(\frac{|\sigma_{xy}|}{A}-\sigma_c\right) \frac{\bar\sigma}{\tau_1} \ ,
\end{eqnarray}
where $A$, $\sigma_{xy}$ and $u_x$ are functions of $x$ and $t$ at $y\!=\!0$. Eq. (\ref{friction}) is analogous to an elasto-plastic decomposition $\dot\epsilon^{\rm el}\!=\!\dot\epsilon^{\rm tot}\!-\!\dot\epsilon^{\rm pl}$ ($\epsilon$ is a strain measure), corresponding to elastic, total and plastic strain rates, respectively. The $\theta$-function expresses the fact that plasticity is an intrinsically threshold phenomena, where $\sigma_c$ is the shear strength or yield stress of the microcontacts. The macroscopic stress $\sigma_{xy}$ is enhanced at the microcontacts level by a factor $A^{-1}\!\!\gg\!\!1$. $\mu_0$ is the interfacial elastic stiffness, which is usually not included in friction models (but note \cite{10SNBZ}), $h$ is the effective height of the interface and $\tau_1$ is a basic timescale of irreversible interfacial processes. $\tau_1^{-2}\!=\!\tau_2^{-2}\!+\!\left(\pa_t u_x/D\right)^2$, i.e. $\tau_1$ is an incoherent sum of an intrinsic timescale $\tau_2$ and a kinetic time scale $D/\pa_t u_x$, where $D$ is a typical sliding distance \cite{06BC}.

The dynamics of $A$, Eq. (\ref{contact}), consists of two contributions; the first tends to increase $A$ to a limiting value $A_0\!\!\ll\!\!1$ on a timescale $\tau_0$. In general $A_0$ depends on the normal stress and may ``age'' logarithmically on long timescales, but here it is constant. The second contribution accounts for the reduction of $A$ due to irreversible processes (plastic deformation and eventually fracture) and is similar to the corresponding term in Eq. (\ref{friction}). The dimensionless parameter $\kappa$ accounts for the possibility that $A$ decays on a timescale somewhat different than $\bar\sigma$, as suggested in \cite{10BDRF}. Eqs. (\ref{separation}-\ref{friction}) constitute a ``minimal'' continuum friction model based on the detachment and reattachment dynamics of interfacial contacts. A closely related discrete model has been recently proposed in \cite{09BBU}.

Focusing on the limit $\tau_2 \!\!\gg\!\! D/|\pa_t u_x|$ and defining $\tilde t\!\!=\!\!t/\tau_0$, $\tilde x\!\!=\!\!x/D$, $\tilde u\!\!=\!\!u/D$, $\tilde A\!\!=\!\!A/A_0$, $\tilde\sigma_{xy}\!\!=\!\!\sigma_{xy}/\sigma_c A_0$ and $\tilde{\bar\sigma}\!\!=\!\!\bar\sigma/\sigma_c A_0$, Eqs. (\ref{contact}-\ref{friction}) become
\begin{eqnarray}
\label{contact1} \pa_{\tilde t} \tilde A &=& (1-\tilde A) - \theta\left(|\tilde\sigma_{xy}|/\tilde A-1\right) \kappa\, \tilde A\, |\pa_{\tilde t} \tilde u_x| \ ,\\
\label{friction1} \pa_{\tilde t} \tilde{\bar\sigma} &=& \alpha\, \tilde A\, \pa_{\tilde t} \tilde u_x  - \theta\left(|\tilde\sigma_{xy}|/\tilde A-1\right) \tilde{\bar\sigma}\, |\pa_{\tilde t} \tilde u_x| \ ,
\end{eqnarray}
with $\alpha\!=\!\mu_0 D/\sigma_c h$. Finally, we rewrite Eq. (\ref{separation}) as $\tilde \sigma_{xy}\! = \!\tilde{\bar\sigma}\!+\! \tilde\eta \pa_{\tilde t} \tilde u_x$, with $\tilde\eta\!\!=\!\!\eta D/\sigma_c\tau_0A_0$, which completes the derivation of our dimensionless friction law.

Macroscopic frictional phenomena intrinsically involve the coupling between bulk elastic deformation and the dissipative dynamics of the interface.
Therefore, we should couple our friction law to a linear-elastic bulk of height $H$. The bulk is described by Lam\'e equation for the displacement field $\B u(x,y,t)$ \cite{93Lawn}. $\sigma_{xy}$ in Eq. (\ref{separation}) is a boundary condition at $y\!=\!0$ (recall that $u_y(x,0,t)\!=\!0$). In the limit of small $H$, $\sigma_{xy}$ is no longer a boundary condition, but rather a term in a 1D equation \cite{supplementary}
\begin{equation}
\label{elastic}
\tilde\rho \,\pa_{\tilde t \tilde t} \tilde u_x = \tilde\mu\, \pa_{\tilde x \tilde x} \tilde u_x + \tilde \sigma_d - \tilde{\bar\sigma} - \tilde\eta\, \pa_{\tilde t} \tilde u_x \ .
\end{equation}
Here $\tilde\rho\!=\!\rho H D/\sigma_c \tau_0^2 A_0$ and $\tilde\mu\!\!=\!\!\mu H/\sigma_cD A_0$, where $\rho$ is the mass density and $\mu$ is the bulk shear modulus \cite{supplementary}. $\tilde\sigma_d$ is the dimensionless external driving stress.

In order to mimic the experimental edge-loading setup of \cite{04RCF, 07RCF, 10BDRF, 10BDCF}, we study Eqs. (\ref{contact1}-\ref{elastic}) for $\tilde x\!\ge\!0$ with a localized driving force of the form $\tilde \sigma_d(\tilde x,\tilde t)\!=\!\tilde K_d \left[\tilde v_d \tilde t \!-\! \tilde u_x(0,\tilde t) \right]\! \delta(\tilde x)$. Here $\tilde K_d$ and $\tilde v_d$ are the rescaled driving spring constant and velocity, respectively, and $\delta(\cdot)$ is a $\delta$-function.

\begin{figure}[here]
\centering \epsfig{width=0.5\textwidth,file=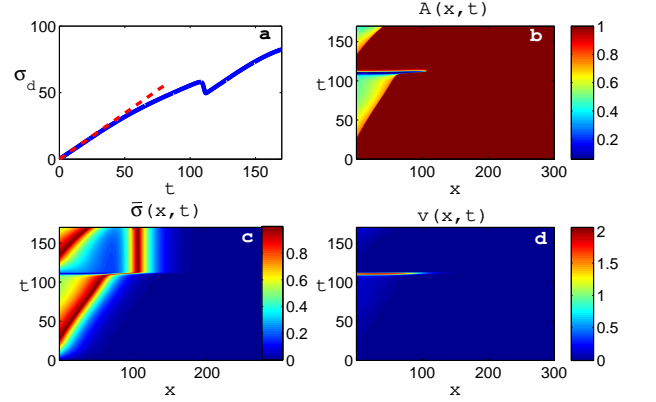}
\caption{(Color online) A solution of Eqs. (\ref{contact1}-\ref{elastic}) with $\tilde \rho\!=\!0.01$, $\tilde \mu\!=\!300$, $\kappa\!=\!8$, $\alpha\!=\!1.05$, $\tilde \eta\!=\!0.1$, $\tilde K_d\!=\!2.5$ and $\tilde v_d\!=\!0.3$. (a) The driving force $\tilde \sigma_d(t)$. (b) The contact area $\tilde A(x,t)$. (c) $\tilde{\bar\sigma}(\tilde x,\tilde t)$. (d) The slip velocity $\tilde v(\tilde x,\tilde t)\!=\!\pa_{\tilde t} \tilde u_x(\tilde x,\tilde t)$. For simplicity, the tilde's are omitted from the axis labels.}\label{fig1}
\end{figure}

We first studied our model numerically \cite{comment}. A typical solution is presented in Fig. \ref{fig1}. Initially, the response is purely elastic, i.e. $\tilde A\!=\!1$ in panel (b) and the driving force rises almost linearly in panel (a). At $\tilde t^*$ ($\simeq\!25$ here), a cracklike front is initiated at the trailing edge ($\tilde x\!=\!0$), leaving behind a reduced contact area $\tilde A\!<\!1$. This cracklike front is accompanied by a relatively low slip rate, see panel (d), but by a significant stress transfer resulting in an inhomogeneous stress state, see panel (c). Finally, this front induces a small, but noticeable, drop in the driving force, see the early time deviation from the straight dashed line in panel (a).

Later, at $\tilde{t}\,\hat{}$ ($\simeq\!110$), an instability in which a short-lived cracklike front is initiated at the trailing edge and arrests at a finite distance (smaller than the system's length $\tilde L$) occurs. The front's velocity is much higher than the one initiated at $\tilde t^*$ (compare the slopes in panel (b) and recall that the smaller the slope, the higher the velocity) and is accompanied by a sharp increase in the slip rate, see panel (d). Finally, the transient front transfers a highly concentrated stress distribution to the interior of the material, see panel (c), and the jump in slip rate at the edge causes an abrupt drop in the driving force, see panel (a).

After the transient front arrests, the contact area recovers to its background level (in general this will not be the case since normal stress variations, which are absent here, will carry memory of past deformation), see panel (b), but a highly concentrated stress distribution remains inside the material, as a signature of the irreversible slip that took place, see panel (c). Another front, similar to the one initiated at $\tilde t^*$, is initiated at the trailing edge soon after ($\tilde t\!\simeq\!130$). It eventually triggers another instability that results in a transient front that propagates deeper into the material (not shown). This complex precursory activity repeats itself until a front reaches the leading edge ($\tilde x\!=\!\tilde L$), whereupon macroscopic sliding occurs (not shown), in agreement with experimental observations \cite{07RCF}. We focus on the first precursory event.

In order to develop a theoretical understanding of the cracklike dynamics that precedes the onset of sliding, we first focus on the elastic response at $\tilde t\!<\!\tilde t^*$. In this regime, the contact area remains intact, $\tilde A\!=\!1$, and the $\theta$-function in Eqs. (\ref{contact1}-\ref{friction1}) vanishes. Then
\begin{equation}
\label{elastic_solution}
\tilde u_x(\tilde x,\tilde t) = \frac{\tilde K_d \,\tilde v_d \,\tilde t}{\tilde K_d+\sqrt{\alpha \tilde \mu}} \exp\left(-\sqrt{\frac{\alpha}{\tilde \mu}}\,\tilde x\right) \theta(\tilde x)\quad\hbox{for}\quad \tilde t<\tilde t^* \ ,
\end{equation}
where the inertial term, which is negligible in this regime, was omitted. This solution is valid until $\tilde{\bar\sigma}/\tilde A$ first equals unity (note that $\tilde \eta\pa_{\tilde t} \tilde u_x$ is negligible in this regime, thus we can replace $\sigma_{xy}$ with $\bar\sigma$). This happens when $\tilde{\bar\sigma}(0,\tilde t^*)\!=\!\alpha \tilde u_x(0,\tilde t^*)\!=\!1$, i.e. $\tilde t^*\!=\!(\tilde K_d+\sqrt{\alpha \tilde \mu})/(\alpha \tilde K_d \tilde v_d)$. The exponential spatial variation in Eq. (\ref{elastic_solution}) is the 1D Green's function (whose analog at higher dimensions is a power-law distribution) and hence it characterizes any purely elastic region, even at times $\tilde t\!>\!\tilde t^*$.

We proceed to discuss the cracklike front that initiates at $\tilde t^*$. To understand its nature, we adopt a fracture mechanics perspective \cite{93Lawn}, which tells us that our edge-loading system is intrinsically {\em stable}. That is, the stress level near the tip of a crack is a {\em decreasing} function of the length of the crack under constant loading at the trailing edge. That means that the tip region, which has to satisfy some fracture criterion (in our case $\tilde \sigma_{xy}/\tilde A\!=\!1$), cannot propagate unless the external loading is increased. This implies that the dimensionless velocity $\bar c$ of such a crack must satisfy $\bar{c} \propto \tilde v_d$. Our system is actually the frictional analog of the famous Obreimoff's tensile fracture experiment \cite{93Lawn}. The prediction $\bar{c} \propto \tilde v_d$ is fully supported by the numerical solutions. The fact that fracture mechanics predicts a stable front is important since it immediately suggests that the instability taking place at $\tilde t\,\hat{}$  has a frictional origin, absent in tensile fracture.

The front that initiates at $\tilde t^*$ plays an important role in triggering the frictional instability at $\tilde t\,\hat{}$. In particular, it is responsible for the transfer of stress from the trailing edge to the interior of the material and hence for the buildup of an inhomogeneous stress distribution. Such inhomogeneous stress distributions were shown to play an important role in selecting various interfacial cracklike modes in \cite{10BDCF}. The frictional instability at $\tilde t\,\hat{}$ is accompanied by a transient cracklike front whose dimensionless velocity $\tilde c \!\gg\!\bar c$. Understanding the nature and properties of this front is a major goal of this Letter.

\begin{figure}[here]
\centering \epsfig{width=0.52\textwidth,file=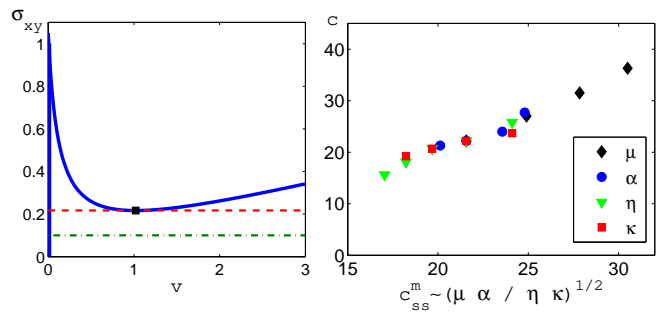}
\caption{(Color online) Left: The steady friction law $\tilde\sigma_{xy}(\tilde v)$ (parameters as in Fig. \ref{fig1}) (solid line). The horizontal lines correspond to homogeneous loading. Right: The transient front velocity $\tilde c$ vs. the minimal steady state velocity $\tilde{c}_{ss}^m\!\sim\!\sqrt{\tilde \mu \alpha/\tilde \eta \kappa}$. The tilde's are omitted from all quantities.} \label{fig2}
\end{figure}

We do not consider here the onset (nucleation) of instability, which we observed to involve fast, Rayleigh-like, contact area disturbance initiated at the trailing edge, but rather focus on the emerging short-lived cracklike fronts. In order to develop a theoretical understanding of these fronts we should discuss some basic properties of our friction law. Consider the spatially homogeneous fixed-points of these equations as a function of the slip rate $\tilde v\!=\pa_{\tilde t} \tilde u_x$. At $\tilde v\!=\!0$ there exist only elastic solutions with $0\!\le\!\tilde\sigma_{xy}\!<\!1$. For $\tilde v\!>\!0$, we find the following sliding solutions
\begin{equation}
\label{sliding}
\tilde\sigma_{xy}(\tilde v) = \frac{\alpha}{1+\kappa \,\tilde v} + \tilde \eta \,\tilde v \quad\hbox{for}\quad \tilde v > 0 \ .
\end{equation}
Since $\tilde \sigma_{xy}\!\ge\!\tilde A$ for sliding solutions (the $\theta$-function equals unity), internal consistency demands that $\alpha\!\ge\!1$. An important feature of the sliding friction law in Eq. (\ref{sliding}) is its nonmonotonic behavior stressed above, where $\tilde \sigma_{xy}(\tilde v)$ attains a minimum at $\tilde v_m$, with a finite value $\tilde \sigma_d^c\!=\!\tilde\sigma_{xy}(\tilde v_m)$. The complete steady state sliding friction law is shown in Fig. \ref{fig2} (left). The parameters $\alpha$ and $\kappa$, control the rate of velocity-weakening for $0\!<\tilde v\!\lesssim\!\tilde v_m$, while $\tilde \eta$ controls the rate of velocity-strengthening for $\tilde v\!\gtrsim\!\tilde v_m$. Their relative values determine the crossover at $\tilde v\!=\!\tilde v_m$. We suggest that this nonmonotonic behavior is generic \cite{06BC, 10RL}.

It is instructive to consider also situations in which the homogeneous loading $\tilde \sigma_d$, rather than the slip rate $\tilde v$, is controlled. In this case, for $\tilde \sigma_d\!<\!\tilde\sigma_d^c$ (see the dashed-dotted line in Fig. \ref{fig2}), we have only one solution (with $\tilde v\!=\!0$), which is a stable solution. At $\tilde\sigma_d\!=\!\tilde\sigma_d^c$ (see the dashed line in Fig. \ref{fig2}) another solution emerges at $\tilde v_m$ (marked by a filled square in Fig. \ref{fig2}). This allows for the propagation of a steady state ``phase transition'' front in which a sliding domain invades a jammed one. For $\tilde\sigma_d\!>\!\tilde\sigma_d^c$ we have two finite $\tilde v$ solutions, one with $0\!<\!\tilde v\!<\!\tilde v_m$ in the velocity-weakening (unstable) regime and one with $\tilde v\!>\!\tilde v_m$ in the velocity-strengthening (stable) regime, which also allow for steady state front solutions.

Can one relate the steady state homogeneously loaded fronts to the transient fronts observed under inhomogeneous loading?
Our basic idea is that {\em the transient fronts are short-lived excitations of fronts corresponding to the {\bf \em minimum} of the sliding friction law}.
To test this idea, denote the minimal velocity of steady state fronts by $\tilde c_{ss}^m$, set $\pa_{\tilde t}\!=\!-\tilde c_{ss}^m \pa_{\tilde x}$ in Eqs. (\ref{contact1}-\ref{elastic}) and note that Eq. (\ref{sliding}) implies $\tilde v_m\!\sim\!\sqrt{\alpha/\tilde\eta\kappa}$ and $\tilde\sigma_d^c\!\sim\!\tilde\eta \tilde v_m$, for $\kappa\!\!\gg\!\!1$. Hence, the scaling version of Eqs. (\ref{friction1}) is $\tilde c_{ss}^m/\tilde\ell_{\bar\sigma} \sim \tilde v_m$ and of Eq. (\ref{elastic}):
\begin{eqnarray}
\label{scaling}
\!\!\!\! \tilde\mu \,\tilde v_m/\tilde \ell_{\bar\sigma} \,\tilde c_{ss}^m + \tilde \sigma^c_d - \tilde{\bar\sigma} \!-\! \tilde\eta\, \tilde v_m\!=\!0 ~\Longrightarrow~ \tilde \mu\, \tilde v_m \sim \tilde\ell_{\bar\sigma}\, \tilde c_{ss}^m \ .
\end{eqnarray}
Here $\tilde \ell_{\bar\sigma}$ is spatial scale of variation of $\tilde{\bar\sigma}$, $\tilde \sigma_d^c\!\sim\!\tilde \eta \tilde v_m$ and $\tilde{\bar\sigma}\!\simeq\!1$ (near the front edge, where the $\theta$-function first equals unity) were used and inertia was neglected. Solving for $\tilde c_{ss}^m$ and $\tilde \ell_{\bar\sigma}$, we obtain (with $\tilde H\!=\!H/D$)
\begin{eqnarray}
\label{speed}
\tilde c_{ss}^m/\tilde s\sim \sqrt{\tilde H/\tilde \ell_\kappa} \quad\Longrightarrow\quad \tilde c_{ss}^m \sim \sqrt{\tilde \mu \alpha/\tilde\eta \kappa}
\end{eqnarray}
and $\tilde\ell_{\bar\sigma}\!\sim\!\sqrt{\tilde H \tilde \ell}$. $\tilde \ell\!=\!\tilde \mu/\tilde H$ and $\tilde \ell_\kappa\!=\!\tilde\mu\kappa/\tilde H\tilde\eta \alpha\!>\!\tilde \ell$ are two rescaled friction-related lengthscales, and $\tilde s\!=\!\tilde \mu/\tilde H\tilde \eta$.

Eq. (\ref{speed}) contains an elastic bulk property $\tilde\mu$ and the three friction parameters $\alpha, \tilde\eta, \kappa$. Adopting our idea that $\tilde c\!\sim\!\tilde c_{ss}^m$, it provides an analytic prediction for the transient front velocity $\tilde c$. We measured $\tilde c$ directly from numerical solutions. In Fig. \ref{fig2} (right) we show $\tilde c$ vs. the minimal steady state velocity $\tilde{c}_{ss}^m\!\sim\!\sqrt{\tilde \mu \alpha/\tilde \eta \kappa}$. Different symbols denote variations of each parameter, shown in the legend, while the other three were held fixed. The result is striking: the data points collapse on a single linear curve with a slope of order unity. This result provides strong evidence in favor of our basic idea; the properties of the transient inhomogeneously loaded fronts are indeed determined by the steady state front solutions at the minimum of the sliding friction law.

A crucial point to note is that Eq. (\ref{speed}) suggests that $\tilde c$ is {\em independent} of the sound speed, $\tilde c_s\!=\!\sqrt{\tilde\mu/\tilde\rho}$. For example, in Fig. \ref{fig1}, $\tilde c\!\simeq\!20$ while $\tilde c_s\!\simeq\!170$, i.e. $\tilde c$ and the sound speed are separated by an order of magnitude. Instead of $\tilde\rho$ a combination of friction parameters determine $\tilde c$. In particular, we highlight the fact that $\tilde c$ depends explicitly on the interfacial stiffness $\mu_0\!\sim\!\alpha$ \cite{10SNBZ}. It is important to note that the spectrum of front solutions starts at a {\bf \em finite} value $\tilde c_{ss}^m$ and hence there exists a ``forbidden gap'' of velocities below it, in contrast to ordinary tensile cracks (or standard ``phase transition fronts''). This is consistent with the experiments in \cite{04RCF, 10BDCF}, where the slow front is always characterized by a well-defined finite velocity. Therefore, we propose that the transient fronts observed here are analogous to the slow fronts discovered recently \cite{04RCF, 10BDCF} and whose velocity should be described by $c_{ss}^m$.

What about the sub-Rayleigh/super-shear fronts? We suspect that this is a matter of the mechanical conditions at nucleation. When the stored energy levels are low, as in the examples considered above, $\tilde c_{ss}^m$ may be the most relevant ``attractor'' for the dynamics. On the other hand, when the stored energy levels at nucleation are higher, a continuous spectrum of solutions with $\tilde c\!>\!\tilde c_{ss}^m$ may be excited. This seems perfectly consistent with the findings of \cite{08SBZN, 10BDCF}, where mode selection is shown to be controlled by the conditions at nucleation.

While we have not yet numerically studied our model in 2D and hence at the moment cannot comment on the validity of our main idea in 2D, we can show that $\tilde c_{ss}^m$ varies smoothly from 1D to 2D as $\tilde H$ is increased and hence it is qualitatively independent of dimensionality. To see this we note that the 1D limit is valid for $\tilde H\!\ll\!\tilde\ell_\kappa, \tilde\ell$. Consider then the 2D regime $\tilde\ell\!\ll\!\tilde H$. The distinguishing feature of dimensions higher than 1 is the existence of (cracklike) power-law singularity at $\tilde\ell\!\ll\!|\tilde x|\!\ll\!\tilde H$. One can show that in the singular region the boundary conditions of the present model reduce to those of \cite{02BM, 05BMM}, who found that $\tilde v\!\sim\!|\tilde x|^{(-1+\epsilon)/2}$, where $\tan\!\left(\pi\epsilon/2\right)\!\sim\! \tilde c_{ss}^m/\tilde s$.

If we assume $\tilde H\!\!\gg\!\!\tilde\ell_\kappa,\!\tilde\ell$, then our result is identical to Eq. (11) in \cite{02BM}, which here reads $\epsilon\!\!\sim\!\!\ln\!\left(\!\tilde H\!/\!\tilde\ell_\kappa\!\right)\!/\!\ln\!\left(\!\tilde H\!/\!\tilde\ell\right)$. If we assume $\tilde\ell\!\!\ll\!\!\tilde H\!\!\ll\!\!\tilde\ell_\kappa$, we can still use the relation $\tilde c_{ss}^m/\tilde\ell_{\bar\sigma} \!\sim \!\tilde v_m$, but now with $\tilde\ell_{\bar\sigma}$ replaced by $\tilde\ell$ and $\tilde v_m$ by $\tilde v_m (\tilde H/\tilde \ell)^{(1-\epsilon)/2}$, where the latter describes the increase of the slip rate on a scale $\tilde\ell$ (it decays back to $\tilde v_m$ on a scale $\tilde H\!\!\gg\!\!\tilde \ell$). The resulting expression is $\tilde c_{ss}^m/\tilde s \!\sim\! \sqrt{\tilde H/\tilde\ell_\kappa}(\tilde\ell/\tilde H)^{\epsilon/2}$. It matches the 1D result of Eq. (\ref{speed}), when $\tilde H\!\sim\!\tilde\ell$, and the 2D result stated above, when $\tilde H\!\sim\!\tilde\ell_\kappa$. We reiterate that the most important implication of this 2D scaling analysis is that the presence of cracklike singularity does not change the qualitative properties of the spectrum of steady state front velocities, which still starts at a finite value.

We believe our results derive from robust properties of dry friction and hence are general (and not model-specific). In future work we plan to test our ideas in 2D numerical simulations, to include the effect of normal stress variations and aging on the contact area, to address the full spectrum of cracklike fronts and to make quantitative comparison to experimental data.

\begin{acknowledgments}
We thank O. Ben-David and J. Fineberg for numerous insightful discussions. EB acknowledges support of the James
S. McDonnell Foundation, the Harold Perlman Family Foundation and the Robert Rees Applied Research Fund. EAB acknowledges support of the Erna and Jacob Michael visiting professorship funds at Weizmann Institute of Science.
MU acknowledges support by the Israel Science Foundation, Grant 1109/09 and by the German-Israeli Project Cooperation Program (DIP).

\end{acknowledgments}


\begin{thebibliography}{99}
\bibitem{comment} We used $\sigma_{xy}\!\simeq\!\bar\sigma$ in the $\theta$-function, which has a minor effect on the results.
\end{thebibliography}
\end{document}